\documentclass[pra,twocolumn,showpacs,preprintnumbers,superscriptaddress]
{revtex4-2}

\usepackage{times}
\usepackage{bm}
\usepackage{float}
\usepackage{graphicx}
\usepackage{amsbsy}
\usepackage{amsmath,mathtools}
\usepackage{amsfonts}
\usepackage{amsthm}
\usepackage{xcolor}

\begin{document}
\theoremstyle{plain}
\newtheorem{theorem}{Theorem}
\newtheorem{lemma}[theorem]{Lemma}
\newtheorem{corollary}[theorem]{Corollary}
\newtheorem{proposition}[theorem]{Proposition}
\newtheorem{conjecture}[theorem]{Conjecture}
\theoremstyle{definition}
\newtheorem{definition}[theorem]{Definition}
\theoremstyle{remark}
\newtheorem*{remark}{Remark}
\newtheorem{example}{Example}

\title{Search for an efficient entanglement witness operator for bound entangled states in bipartite quantum systems}
\author{Shruti Aggarwal, Satyabrata Adhikari}
\email{shruti_phd2k19@dtu.ac.in, satyabrata@dtu.ac.in} \affiliation{Delhi Technological University, Delhi-110042, Delhi, India}

\begin{abstract}
Entanglement detection problem is one of the important problem in quantum information theory. Gurvit \cite{gurvit} showed that this problem is NP complete and thus this may be the possible reason that only one criterion is not sufficient to detect all entangled states. There are some powerful entanglement detection criterion such as partial transposition criterion, realignment criterion but it may not be possible to implement them successfully in the experiment. This situation can be avoided if the entanglement is detected through the construction of witness operator method. In this work, we take an analytical approach to construct a witness operator. To achieve this task, we first construct a linear map using partial transposition and realignment operation. Then we find some conditions on the parameters of the map for which the map represent a positive map. Further, we have constructed a Choi matrix corresponding to the map and have shown that it is not completely positive. We then construct an entanglement witness operator, which is based on the linear combination of the function of Choi matrix and the identity matrix and it can detect both NPTES and PPTES. Finally, we prove its efficiency by detecting several bipartite bound entangled states which were previously undetected by some well-known separability criteria. We also compared the detection power of our witness operator with three well-known powerful entanglement detection criteria, namely, dV criterion, CCNR criterion and the separability criteria based on correlation tensor (CT) proposed by Sarbicki et. al. and find that our witness operator detect more entangled states than these criterion.
\end{abstract}
 \maketitle

\section{Introduction}
There are certain quantum information processing tasks such as quantum teleportation \cite{Bennett1}, superdense coding \cite{Bennett2} that may be implemented in an efficient way if we use entangled states as a resource state in comparison to the optimal classical prepare and measure strategy \cite{popescu}. Entanglement also plays a vital role in quantum cryptography \cite{Gisin} and quantum computation \cite{jozsa}. Ekert and Jozsa found that the non-local correlation that exist in the entangled state is responsible for the computational speed-up in the known quantum algorithms \cite{ekert100}. Thus, entanglement \cite{horodecki4} serve as a basic ingredient in quantum information theory \cite{nielsen}.\\
A state in the tensor product $M_m \otimes M_n$ of matrix algebras is called separable if it is a convex combination of pure product states, which are rank one projections onto product vectors. Non-separable states are called entangled states. Since entanglement is a crucial prerequisite to perform any quantum communication task so its detection is equally important. It is in general difficult to determine whether a given state is entangled or separable. In the low dimensional system e.g. in $2\otimes 2$ and $2 \otimes 3$ dimensional system, the partial transposition method works very efficiently for the detection of entangled states. This entanglement detection criterion was first noted by Peres and Horodecki (PH) \cite{ph,horo96}. It is also known as PH criterion and it involves partial transposition operation. Partial transposition means the transposition operation performed on one of the subsystem of the composite system. Mathematically, the partial transpose $(x \otimes y)^\tau$ of $x \otimes y \in M_m \otimes M_n$ is given by $x^t \otimes y$ with the usual transpose $x^t$. If we identify $M_m \otimes M_n$ with the block matrices $M_m(M_n)$, then the partial transpose corresponds to the block-wise transpose. PH criterion provides the necessary and sufficient condition for the detection of entangled states in $2\otimes 2$ and $2 \otimes 3$ dimensional system. The criteria states that a quantum state is separable if and only if the eigenspectrum  of partial transposed state contain positive eigenvalues.\\
Although there is a lot of progress in this line of research but till now there does not exist any necessary and sufficient condition for entanglement detection in multipartite and higher dimensional bipartite system. This may be due to the fact that in higher dimensional systems, the eigenspectrum  of partial transposition of entangled state contain positive eigenvalues. This type of states are known as positive partial transpose entangled states (PPTES) or bound entangled states (BES). Since PPTES are not detected by partial transposition method so other criterion such as the computable cross norm and realignment (CCNR) Criterion \cite{rudolf,kchen}, range criterion \cite{phorodecki}, majorization criterion \cite{nielsenkempe}, covariance matrix criterion \cite{guhnehyllus} developed in detecting the PPTES. In spite of these criterion, there exist other approaches such as construction of positive but not completely positive map, which may also help in the detection of entangled states. The well known example of such map is the Reduction map \cite{mphorodecki}. Since entanglement detection problem is a NP hard problem so all entangled states cannot be detected by just one criterion and thus numerous criterion has been developed for the detection of entanglement.\\
In this work, we employ the theoretical approach based on the linear maps which are positive but not completely positive, also referred to as PnCP maps. Positive linear maps are useful in the entanglement detection \cite{bmt}. The simplest example of a PnCP map is the partial transposition (PT) map. Any positive, but not completely positive map give rise to a construction of entanglement witnesses via the Choi-Jamiolkowski isomorphism \cite{jamio}. Entanglement witness operator plays a significant role in the entanglement detection problem since if we have some prior partial information about the state which is to be detected then entanglement can be detected in the experiment by the construction of witness operator \cite{guhne2}. Mathematically, entanglement witness operator is defined as a Hermitian operator $W$ that satisfies the following properties: \begin{eqnarray}
&&(i) Tr(W\sigma)\geq 0,~~ \textrm{for all separable state}~~ \sigma~~ \textrm{ and}
\nonumber\\&& (ii) Tr(W\rho)< 0,~~\textrm{for at least one entangled state}~\rho\nonumber
\label{defwit}
\end{eqnarray}
The first example of witness operator is the Bell operator, the operator which appear in the Bell-CHSH inequality \cite{bellchsh}. There are two types of witness operator: (i) Decomposable witness operator and (ii) Non-decomposable witness operator. Decomposable witness operator can only detect negative partial transpose entangled states (NPTES) while the latter one detect NPTES as well as PPTES \cite{lewenstein}. There are lot of works in the literature that dealt with the entanglement detection problem using witness operator as a tool \cite{ng,sha,ak}. One can also see these nice reviews that included the topic of entanglement witnesses \cite{guhne1,xzhang}.\\
This paper is organized as follows: In section-II, some basic definitions and results are stated to be used in the later sections. In section III, we construct a linear map which is shown to be positive but not completely positive. In section IV, we have constructed a Hermitian operator using the Choi matrix generated from the constructed linear map and found in the later stage that the constructed Hermitian operator satisfies all the properties of a witness operator. We also show that the constructed witness operator detect PPTES. In section V, we show that our witness operator is more efficient than the recently proposed strong criterion in the sense that our witness operator detect more entangled states than the existing criterion.

\section{Preliminaries}
\subsection{Basic Definitions}
\definition \textit{Positive map:} A linear map $\phi : A\rightarrow B$ is called positive map if it maps positive elements of $A$ to positive elements of $B$.
\definition \textit{Completely Positive map:} Let $\phi : A\rightarrow B$ be a linear map, and let $k \in \mathbb{N}$. the set of natural numbers. Then $\phi$ is \textit{k}-positive if $\phi \otimes i_k : A \otimes M_k \rightarrow B \otimes M_k$ is positive, $i_k$ denotes the
identity map on $M_k$. $\phi$ is said to be completely positive if $\phi$ is $k$-positive $\forall$ $k \in \mathbb{N}$.

\definition \textit{Choi map:} Let $\Phi: M_{d_1}(\mathbb{C}) \rightarrow M_{d_2}(\mathbb{C})$ be a linear map. The choi matrix for $\Phi$ is defined by the operator
\begin{equation}
C_{\Phi} = \sum_{i,j}^{d_1,d_2} e_{ij} \otimes \Phi(e_{ij}) \label{choi}
\end{equation}
where $\{e_{ij}\}_{i,j=1}^{n}$ denotes the matrix units.


\definition \textit{Realignment operation}\cite{rudolf,kchen}: Let $H= {H_{A}^{d}} \otimes {H_{B}^{d}}$ be a bipartite Hilbert space and the realignment operation R defined by
\begin{equation}
\langle kl|U^R| mn \rangle = \langle km|U|ln \rangle
\label{realignment}
\end{equation}
where $U$ is a unitary operator in the composite Hilbert space $H$. Then for any two matrices $X$ and $Y$, $(X\otimes Y)^R = |X\rangle \langle Y^*|$ where $|X\rangle$ is the row vectorization of $X$ and $*$ is the complex conjugation.

\definition \textit{Correlation matrix}\cite{sarbicki}: For a bipartite state $\rho$ in $\mathcal{H}=\mathcal{H}_A \otimes \mathcal{H}_B$ with dimensions $d_A$ and $d_B$ respectively, the correlation matrix is defined as
\begin{equation}
C_{a,b} = Tr(\rho G_{a}^A \otimes G_{b}^B) \label{cor}
\end{equation}
where $G_{a}^A$ and $G_{b}^B$ denote arbitrary orthonormal basis in $\mathcal{B(H_A)}$ and $\mathcal{B(H_B)}$. In particular, if $(G_{a}^A)_{a\neq 0}$, $(G_{b}^B)_{b\neq 0}$ represent orthonormal traceless operators and $G_{0}^A=\frac{I_{A}}{\sqrt{d_{A}}}$, $G_{0}^B=\frac{I_{B}}{\sqrt{d_{B}}}$ then $\frac{1}{\sqrt{d_{A}}}G_{a}^A \in \mathcal{B(H_A)}$ and $\frac{1}{\sqrt{d_{B}}}G_{b}^B \in \mathcal{B(H_B)}$ forms a canonical basis.

\subsection{A few Results}
\noindent \textbf{Result-1 \cite{weyl}}: Let $X$ and $Y$ be hermitian matrices in $M_n(\mathbb{C})$, then
\begin{equation}
\lambda_{min}(X) + \lambda_{min}(Y) \leq \lambda_{min}(X+Y)
\label{weyl}
\end{equation}
\textbf{Result-2 \cite{ieee}}: For any two hermitian matrices $A$ and $B$ in $M_n(\mathbb{C})$, we have
\begin{equation}
\lambda_{min}(A)Tr[B]\leq Tr[AB] \leq \lambda_{max}(A) Tr[B]  \label{res2}
\end{equation} where $\lambda_{min}$ and $\lambda_{max}$ are the minimum and maximum eigenvalues of $A$.\\
\textbf{Result-3 \cite{john}}:
Let $\sigma =
\begin{bmatrix}
X & Y \\
Y^{\dagger} & Z \\
\end{bmatrix}$ be the density matrix. Then $\sigma$ is separable if
\begin{equation}
||Y||_2^{2} \leq \lambda_{min}(X) \lambda_{min}(Z) \label{res1}
\end{equation}
where $||Y||_{2}^{2}=Tr[YY^{\dagger}]$ denotes the Frobenius norm; $\lambda_{min}(X)$ and $\lambda_{min}(Z)$ denoting the minimum eigenvalues of the matrices $X$ and $Z$ respectively.\\

\textbf{Result-4 \cite{collins}}: Let $\Phi: M_{d_1}(\mathbb{C}) \rightarrow M_{d_2}(\mathbb{C})$ be a linear map and $C_{\Phi}$ be the choi matrix of $\Phi$.  Then the following conditions are equivalent.
\begin{itemize}
\item[1.] $C_{\Phi}$ is positive semi-definite.
\item[2.] $\Phi$ is completely positive.
\item[3.] $\Phi(A)= \sum_{i=1}^{k} V^{*}_i A V_i$ with $V_i:M_{d_2}(\mathbb{C}) \rightarrow M_{d_1}(\mathbb{C})$, a linear map and $k \leq min\{d_1, d_2\}$
\end{itemize} \label{thmchoi}

\subsection{Few Separability Criteria}
\subsubsection{CCNR or Realignment Criterion}
If $\rho$ is separable, then CCNR criterion \cite{rudolph1} gives the following bound
\begin{equation}
||C_{a,b}||_1  \leq 1
\end{equation}
where $C_{a,b}$ denote the correlation matrix defined in (\ref{cor}) and $||A||_1$ denote the trace norm of $A$ and it is defined as $||A||_1= Tr \sqrt{A A^{\dagger}}$.
Furthermore, the realignment criterion state that if $\rho$ is separable, then $||\rho^R||_1 \leq 1$ where $R$ is the realignment operation defined in (\ref{realignment}).

\subsubsection{dV Criterion}
de Vicente (dV) criterion states that if the state $\rho$ in $d_{A}\otimes d_{B}$ dimensional system is separable then the correlation matrix $C_{a,b}$ defined in (\ref{cor}) for the state $\rho$ satisfies the inequality \cite{dv}
\begin{equation}
||C_{a,b}||_1 \leq \frac{\sqrt{d_A d_B (d_A -1)(d_B - 1)}}{2}
\end{equation}

\subsubsection{Correlation Tensor (CT) Criterion}
CT criterion \cite{sarbicki} states that if $\rho$ is separable, then
\begin{equation}
||D_{x}^A C^{can} D_{y}^B||_1 \leq \mathcal{N}_A (x) \mathcal{N}_B (y) \label{ct}
\end{equation}
where $D_x =$ diag($x, 1, 1, . . ., 1$), $D_y =$ diag($y, 1, 1, . . ., 1$),
$\mathcal{N}_A (x)= \sqrt{\frac{d_A - 1 + x^2}{d_A}}$, $\mathcal{N}_B (y)= \sqrt{\frac{d_B - 1 + y^2}{d_B}}$, $x,y \geq 0$, and $C^{can}$ is the correlation matrix defined by canonical basis.

\section{Construction of a map}
\noindent In this section, we first construct a map and then derive the conditions for which the map is positive. Further, we will examine whether the constructed map is completely positive. To fulfill this goal, we provide the form of the Choi matrix generated from the map, which will be shown later that generated Choi matrix is not positive semi-definite. This fact shows that the constructed map is positive but not completely positive. \\
To start with, let us consider a map $\Phi_{\alpha,\beta}: M_{n}{(\mathbb{C})} \rightarrow M_{n}{(\mathbb{C})} $, $n\geq2$ defined as
\begin{equation}
\Phi_{\alpha,\beta}(A) = \alpha  A^{T_B} + \beta  A^R
\label{phi}
\end{equation}
where $A \in M_n{(\mathbb{C})}$, $\alpha,\beta \in \mathbb{R^{+}}$. $R$ represents the realignment operation and $T_B$ denote the partial transposition operation with respect to subsystem $B$. The parameters $\alpha$ and $\beta$ is chosen in such a way that the map $\Phi$ represent a positive map and then we fix the parameters $\alpha$ and $\beta$.\\
Now our prime task is to search for the parameters $\alpha$ and $\beta$ for which the map $\Phi_{\alpha,\beta}$ will be positive. Once chosen the parameters $\alpha$ and $\beta$, we then move on to investigate whether the positive map will be completely positive for such $\alpha$ and $\beta$.

\subsection{Positivity of the map $\Phi_{\alpha,\beta}$}
Let $A \in M_n{(\mathbb{C})}$ be a positive semi-definite matrix. If $\Phi_{\alpha,\beta}(A)$ is positive for some $\alpha, \beta \in \mathbb{R^{+}}$ then we can say that the map $\Phi_{\alpha,\beta}$ represent a positive map. To show the positivity of the map, it is enough to show $\lambda_{min}(\Phi_{\alpha,\beta}(A)) \geq 0$ for $A\geq 0$. Therefore, $\Phi_{\alpha,\beta}$ will become a positive map if for some fixed $\alpha,\beta \in \mathbb{R^{+}}$, we have
\begin{equation}
\lambda_{min}(\alpha  A^{T_B} + \beta  A^R) \geq 0
\label{mineigen}
\end{equation}
\subsubsection{Determination of the parameters $\alpha$ and $\beta$ for which map $\Phi_{\alpha,\beta}$ represent a positive map}
In this work, we are interested to work with $d_{1} \otimes d_{2}$ dimensional quantum system which can be described by the density operator, say, $\varrho$. Thus, we consider the domain of the map as the set $D$ of all bipartite density matrices $\varrho_{AB} \in M_{d_{1}d_{2}}{(\mathbb{C})}$ such that $\varrho_{AB}^R$ is Hermitian. Therefore, the map $\Phi_{\alpha,\beta}$ given in (\ref{phi}) can be re-expressed as\\
\begin{equation}
\Phi_{\alpha,\beta}(\varrho_{AB}) = \alpha \varrho_{AB}^{T_B} + \beta  \varrho_{AB}^R
\label{phiredefine}
\end{equation}
The map $\Phi_{\alpha,\beta}(\varrho_{AB})$ can be shown to be positive by considering two different cases. In the first case, we assume that the state $\varrho_{AB} \in M_{d_{1}d_{2}}{(\mathbb{C})}$ is PPT and in the second case, we will consider $\varrho_{AB}$ as an NPT entangled state.\\
\textbf{Case-I:} When $\varrho_{AB}$ is a PPT state.\\
Since $\varrho_{AB}$ is a PPT state so $\lambda_{min}(\varrho_{AB}^{T_{B}})\geq 0$, where $\lambda_{min}(.)$ denote the minimum eigenvalue of $(.)$. Thus, using (\ref{phiredefine}) and Weyl's inequality given in (\ref{weyl}), we can write
\begin{equation}
\lambda_{min}(\Phi_{\alpha,\beta}(\varrho_{AB})) \geq \alpha \lambda_{min}(\varrho_{AB}^{T_B}) + \beta \lambda_{min}(\varrho_{AB}^R)
\label{mineig1}
\end{equation}
(i) If $\lambda_{min}(\varrho_{AB}^R)$ is non-negative then $\lambda_{min}\{\Phi_{\alpha,\beta}(\varrho_{AB})\}\geq 0$ for all $\alpha, \beta > 0$. Thus the map $\Phi_{\alpha,\beta}$ is positive $ \forall \alpha, \beta > 0$.\\
(ii) If $\lambda_{min}(\varrho_{AB}^R)$ is negative then we can choose the parameters $\alpha$ and $\beta$ in such a way that $\lambda_{min}\{\Phi_{\alpha,\beta}(\varrho_{AB})\}\geq 0$. Therefore, the chosen parameters $\alpha$ and $\beta$ should satisfy the inequality
\begin{equation}
\frac{\alpha}{\beta} \geq \frac{\lambda_{min}(\varrho_{AB}^R)}{\lambda_{min}(\varrho^{T_B})}
\label{alphabeta1}
\end{equation}
Hence, if $\lambda_{min}(\varrho_{AB}^R)$ is negative then the map  $\Phi_{\alpha,\beta}$ is positive for some parameter $\alpha$ and $\beta$ that satisfies the inequality (\ref{alphabeta1}).\\
\textbf{Case-II:} When $\varrho_{AB}$ is an entangled state.\\
If $\varrho_{AB}$ is an entangled state then $\lambda_{min}(\varrho_{AB}^{T_{B}}) < 0$. Thus, from (\ref{mineig1}), we conclude the following fact:\\
(i) If $\lambda_{min}(\varrho_{AB}^R)$ is non-negative then $\lambda_{min}\{\Phi_{\alpha,\beta}(\varrho_{AB})\}\geq 0$ when the parameter $\alpha, \beta$ satisfy the inequality
\begin{equation}
\frac{\alpha}{\beta} \leq \frac{\lambda_{min}(\varrho_{AB}^R)}{\lambda_{min}(\varrho^{T_B})}
\label{alphabeta2}
\end{equation}
therefore, the map $\Phi_{\alpha,\beta}$ is positive if $\alpha,\beta$ satisfies (\ref{alphabeta2}).\\
(ii) If $\lambda_{min}(\varrho_{AB}^R)$ is negative then the map  $\Phi_{\alpha,\beta}$ never be a positive map for any $\alpha, \beta > 0$.
\subsubsection{Illustration}
\noindent Here, we will analyse the positivity of the map $\Phi_{\alpha,\beta}: M_{d_{1}d_{2}}{(\mathbb{C})} \rightarrow M_{d_{1}d_{2}}{(\mathbb{C})}$ for $d_{1}=d_{2}=2$. For this, we start with a two-qubit with maximally mixed marginals state which is given by \cite{luo}
\begin{equation}
\varrho = \frac{1}{4} [I_2 \otimes I_2 + \sum_{j=1}^{3} t_j \sigma_j \otimes \sigma_j]
\end{equation}
where $I_{2}$ denote the identity matrix of order 2 and $\sigma_{i},i=1,2,3$ denote the Pauli matrices.\\
The matrix representation of $\varrho$ in the computational basis as
\begin{equation}
\varrho=\frac{1}{4}
\begin{pmatrix}
1+t_3 & 0 & 0 & t_1 - t_2 \\
0 & 1-t_3 & t_1+t_2 & 0\\
0 & t_1+t_2 &1-t_3 & 0 \\
t_1-t_2 & 0 & 0 & 1+t_3 \\
\end{pmatrix}
\end{equation}
The eigenvalues of $\varrho$ are non-negative if the following inequality holds \cite{ali}
\begin{eqnarray}
(1 - t_3)^2 &\geq& (t_1 + t_2)^2 \\
(1 + t_3)^2 &\geq& (t_1 - t_2)^2
\label{non-negeigenvalue}
\end{eqnarray}
The matrix representation of partial transposed state $\varrho^{T_{B}}$ is given by
\begin{equation}
\varrho^{T_{B}}=\frac{1}{4}
\begin{pmatrix}
1+t_3 & 0 & 0 & t_1 + t_2 \\
0 & 1-t_3 & t_1-t_2 & 0\\
0 & t_1-t_2 &1-t_3 & 0 \\
t_1+t_2 & 0 & 0 & 1+t_3 \\
\end{pmatrix}
\end{equation}
If $\lambda_{min}(.)$ denote the minimum eigenvalue of $(.)$ then the minimum eigenvalue of $\varrho^{T_{B}}$ is given by
\begin{eqnarray}
\lambda_{min}(\varrho^{T_{B}})&=& min\{\frac{1+t_{1}-t_{2}-t_{3}}{4},\frac{1-t_{1}+t_{2}-t_{3}}{4},\nonumber\\&&\frac{1-t_{1}-t_{2}+t_{3}}{4},
\frac{1+t_{1}+t_{2}+t_{3}}{4}\}
\label{partranseig}
\end{eqnarray}
The realigned matrix $\varrho^{R}$ can be expressed as
\begin{equation}
\varrho^{R}=\frac{1}{4}
\begin{pmatrix}
1+t_3 & 0 & 0 & 1 - t_3 \\
0 & t_1-t_2 & t_1+t_2 & 0\\
0 & t_1+t_2 & t_1-t_2 & 0 \\
1 - t_3 & 0 & 0 & 1+t_3 \\
\end{pmatrix}
\end{equation}
the minimum eigenvalue of $\varrho^{R}$ is given by
\begin{eqnarray}
\lambda_{min}(\varrho^{R})= min\{\frac{1}{2},\frac{t_1}{2},\frac{t_2}{2},\frac{t_3}{2}\}
\label{realigneig}
\end{eqnarray}
We are now in a position to determine the non-negative parameters $\alpha$ and $\beta$ for which the map $\Phi_{\alpha,\beta}$ is positive. To be specific, we discuss here only one parameter family of separable and entangled state.\\
\textbf{Case-I:} One parameter family of separable states $\varrho_{1}$ for which $\lambda_{min}(\varrho_{1}^{R})<0$.\\
Let us take $t_{1}=t_{2}=0$ and $t_{3}=-t, 0< t< 1$. The minimum eigenvalues $\lambda_{min}(\varrho^{T_{B}})$ and $\lambda_{min}(\varrho^{R})$ are given by
\begin{eqnarray}
\lambda_{min}(\varrho_{1}^{T_{B}})= min\{\frac{1+t}{4},\frac{1+t}{4},\frac{1-t}{4},\frac{1-t}{4}\}=\frac{1-t}{4}
\label{min1}
\end{eqnarray}
\begin{eqnarray}
\lambda_{min}(\varrho_{1}^{R})= min\{0,0,\frac{1}{2},\frac{-t}{2}\}=\frac{-t}{2}
\label{min2}
\end{eqnarray}
Since $\lambda_{min}(\varrho_{1}^{T_{B}})>0$ so the state $\varrho_{1}$ represent a family of separable state. Using (\ref{min1}) and (\ref{min2}), we can determine $\alpha$ and $\beta$ for which the map is positive. Therefore, the map $\Phi_{\alpha,\beta}$ will be positive if
\begin{eqnarray}
\frac{\alpha}{\beta} \geq \frac{2t}{1-t}, ~~0<t<1
\label{posmap1}
\end{eqnarray}
\textbf{Case-II:} One parameter family of separable states $\varrho_{2}$ for which $\lambda_{min}(\varrho_{2}^{R})\geq0$.\\
In this case, we take $t_{1}=t_{2}=0$ and $0 \leq t_{3} \leq 1$. In this case, $\lambda_{min}(\varrho_{2}^{T_{B}})$ is found to be same as given in
(\ref{min1}) and $\lambda_{min}(\varrho_{2}^{R})$ is given by
\begin{eqnarray}
\lambda_{min}(\varrho_{2}^{R})= min\{0,0,\frac{1}{2},\frac{t_{3}}{2}\}=0
\label{min4}
\end{eqnarray}
Since $\lambda_{min}(\varrho_{2}^{T_{B}})\geq 0$ so the state $\varrho_{2}$ represent another class of separable states. In this case, it can be easily checked that the map $\Phi_{\alpha,\beta}$ will be positive for all parameters $\alpha, \beta \geq 0$. Thus, we can choose any positive $\alpha$ and $\beta$ to construct a positive map.\\
\textbf{Case-III:} One parameter family of entangled states $\varrho_{3}$ for which $\lambda_{min}(\varrho_{3}^{R})<0$.\\
In this case, we can consider $t_{1}=1$ and $t_{3}=-t_{2}=-t, -1 \leq t \leq 1$. In this case, $\lambda_{min}(\varrho_{3}^{T_{B}})$ and $\lambda_{min}(\varrho_{3}^{R})$ are same and are given by
\begin{eqnarray}
\lambda_{min}(\varrho_{3}^{T_{B}})= \lambda_{min}(\varrho_{3}^{R})= min\{\frac{1}{2},\frac{1}{2},\frac{t}{2},\frac{-t}{2}\}
\label{min5}
\end{eqnarray}
Since the parameter $t$ lying in the interval $-1 \leq t \leq 1$ so it can be easily seen that both $\lambda_{min}(\varrho_{3}^{T_{B}})$ and $\lambda_{min}(\varrho_{3}^{R})$ are negative. Thus, the state $\varrho_{3}$ represent a family of entangled states for $-1 \leq t \leq 1$ and within this interval of $t$, the map $\Phi_{\alpha,\beta}$ can never be positive for any parameters $\alpha, \beta > 0$.\\
\textbf{Case-IV:} An entangled state described by the density operator $\varrho_{4}$ for which $\lambda_{min}(\varrho_{2}^{R})>0$.\\
Let us consider the case when $t_{1}=t_{2}=t_{3}=-1$. $\lambda_{min}(\varrho_{4}^{T_{B}})$ and $\lambda_{min}(\varrho_{4}^{R})$ are given by
\begin{eqnarray}
\lambda_{min}(\varrho_{4}^{T_{B}})= min\{\frac{1}{2},\frac{1}{2},\frac{1}{2},\frac{-1}{2}\}=\frac{-1}{2}
\label{min7}
\end{eqnarray}
\begin{eqnarray}
\lambda_{min}(\varrho_{4}^{R})= min\{\frac{1}{2},\frac{1}{2},\frac{1}{2},\frac{1}{2}\}=\frac{1}{2}
\label{min8}
\end{eqnarray}
 The state $\varrho_{4}$ represent an entangled state. The map $\Phi^{(2,2)}_{\alpha,\beta}$ can never be positive for any parameters $\alpha, \beta > 0$.
 
\subsection{Is the map $\Phi_{\alpha,\beta}$ completely positive?}
\noindent In this subsection, we ask whether the positive map $\Phi^{(d_{1},d_{2})}_{\alpha,\beta}: M_{d_{1}d_{2}}{(\mathbb{C})} \rightarrow M_{d_{1}d_{2}}{(\mathbb{C})}$ for some positive $\alpha,\beta$ is also completely positive. To investigate this question, we consider the domain set as the set of all $2 \otimes 2$ dimensional quantum states. The map for this particular domain will reduce to $\Phi^{(2,2)}_{\alpha,\beta}: M_{4}{(\mathbb{C})} \rightarrow M_{4}{(\mathbb{C})}$. 
From (\ref{phi}), the explicit form of $\Phi_{\alpha, \beta}^{2,2} (A)$ is given as

\begin{eqnarray*}
	\begin{pmatrix}
		(\alpha + \beta) a_{11} && \alpha a_{21} + \beta a_{12} && \alpha a_{13} + \beta a_{21} && \alpha a_{23} + \beta a_{22}\\
		\alpha a_{12} + \beta a_{13} && \alpha a_{22} + \beta a_{14} &&\alpha a_{14} + \beta a_{23}&& (\alpha + \beta) a_{24}\\
		(\alpha + \beta) a_{31} && \alpha a_{41} + \beta a_{32} &&\alpha a_{33} + \beta a_{41} && \alpha a_{43} + \beta a_{42}\\
		\alpha a_{32} + \beta a_{33} && \alpha a_{42} + \beta a_{34} &&\alpha a_{34} + \beta a_{43} &&(\alpha + \beta) a_{44}\\
	\end{pmatrix}
\end{eqnarray*}
where $A= [a_{ij}]_{i,j=1}^{4}$.

We have already shown in the previous section that for this domain set, there exist  $\alpha=\alpha'$ and $\beta=\beta'$ for which the map $\Phi^{(2,2)}_{\alpha',\beta'}$ is positive. The positive map $\Phi^{(2,2)}_{\alpha',\beta'}$ may or may not be completely positive. We will show here that under which condition, the positive map $\Phi^{(2,2)}$ will be completely positive. We have considered here $2 \otimes 2$ dimensional quantum states in the domain to reduce the complexity of the calculation. For the general case, that is,  when the domain of the map $\Phi^{(d_{1},d_{2})}$ contain $d_{1} \otimes d_{2}$ dimensional quantum states, the method of showing the completely positiveness property of the positive map $\Phi^{(d_{1},d_{2})}$ will be same as we have shown below.\\
To start our investigation, we construct the Choi matrix using the result given in (\ref{choi}). The Choi matrix $C_{\Phi^{(2,2)}_{\alpha,\beta}}$ corresponding to the map $\Phi^{(2,2)}_{\alpha,\beta}$ can be constructed as
\begin{equation}
C_{\Phi^{(2,2)}_{\alpha,\beta}} =
\begin{bmatrix}
C_{11} & C_{12}\\
C_{21} & C_{22}\\
\end{bmatrix}
\end{equation}
where the $8\times 8$ block matrices $C_{ij},i,j=1,2$ can be expressed as
\begin{equation}
C_{11} = \begin{pmatrix}
\alpha + \beta & 0 & 0 & 0 & 0 & \beta & 0 & 0\\
0 & 0 & 0 & 0 &\alpha & 0 & 0 & 0 \\
0 & 0 & 0 & 0& 0 & 0 & 0 & 0\\
0 & 0 & 0 & 0& 0 & 0 & 0 & 0 \\
0 & \alpha & \beta & 0& 0 & 0 & 0 & \beta\\
0 & 0 & 0 & 0& 0 & \alpha & 0 & 0 \\
0 & 0 & 0 & 0& 0 & 0 & 0 & 0\\
0 & 0 & 0 & 0& 0 & 0 & 0 & 0\\
\end{pmatrix}
\end{equation}
\begin{equation}
C_{12} = \begin{pmatrix}
0 & 0 & \alpha & 0 & 0 & 0 & 0 & 0\\
\beta & 0 & 0 & 0& 0 &\beta & \alpha & 0\\
0 & 0 & 0 & 0& 0 & 0 & 0 & 0\\
0 & 0 & 0 & 0& 0 & 0 & 0 & 0 \\
0 & 0 & 0 & \alpha & 0 & 0 & 0 & 0\\
0 & 0 & \beta & 0& 0 & 0 & 0 & \alpha+\beta \\
0 & 0 & 0 & 0& 0 & 0 & 0 & 0\\
0 & 0 & 0 & 0& 0 & 0 & 0 & 0\\
\end{pmatrix}
\end{equation}
\begin{equation}
C_{21} = \begin{pmatrix}
0 & 0 & 0 & 0& 0 & 0 & 0 & 0 \\
0 & 0 & 0 & 0& 0 & 0 & 0 & 0 \\
\alpha + \beta & 0 & 0 & 0& 0 & \beta & 0 & 0 \\
0 & 0 & 0 & 0& \alpha & 0 & 0 & 0 \\
0 & 0 & 0 & 0& 0 & 0 & 0 & 0  \\
0 & 0 & 0 & 0& 0 & 0 & 0 & 0  \\
0 & \alpha & \beta & 0& 0 & 0 & 0 & \beta  \\
0 & 0 & 0 & 0& 0 & \alpha & 0 & 0 \\
\end{pmatrix}
\end{equation}
\begin{equation}
C_{22} = \begin{pmatrix}
0 & 0 & 0 & 0& 0 & 0 & 0& 0 \\
0 & 0 & 0 & 0& 0 & 0 & 0& 0 \\
0 & 0 & \alpha & 0& 0 & 0 & 0& 0 \\
\beta & 0 & 0 & 0&0 & \beta & \alpha & 0 \\
0 & 0 & 0 & 0& 0 & 0 & 0& 0 \\
0 & 0 & 0 & 0& 0 & 0 & 0& 0 \\
0 & 0 & 0 & \alpha& 0 & 0 & 0& 0 \\
0 & 0 & \beta & 0& 0 & 0 & 0& \alpha+\beta
\end{pmatrix}
\end{equation}
From the Result-4, it is known that if the Choi matrix $C_{\Phi^{(2,2)}_{\alpha,\beta}}$ represent a positive semi-definite matrix then the map $\Phi^{(2,2)}_{\alpha,\beta}$ is completely positive. Thus, we calculate the eigenvalues $\lambda_{i},i=1,2,...16$ of the Choi matrix $C_{\Phi^{(2,2)}_{\alpha,\beta}}$ and they are given by
\begin{eqnarray}
&&\lambda_{1}=-2\alpha, \lambda_{2}= 2\alpha, \lambda_{3}=2(\alpha+\beta), \nonumber\\&& \lambda_{i}=0,~~i=4,5,...16
\label{eigval}
\end{eqnarray}
(i) When $\alpha=0$ and $\beta\geq 0$, all eigenvalues of $C_{\Phi^{(2,2)}_{\alpha,\beta}}$ comes out to be positive. Thus, $C_{\Phi^{(2,2)}_{\alpha,\beta}}$ will become a positive semi-definite matrix and hence the map $\Phi^{(2,2)}_{\alpha,\beta}$ will become a completely positive map.\\
(ii) The Choi matrix $C_{\Phi^{(2,2)}_{\alpha,\beta}}$ will not represent a positive semi-definite matrix when $\alpha>0$ and $\beta > 0$. Thus, in this case, the map will not be a completely positive map.


\section{Construction of witness operator}
\noindent In this section, we use the Choi matrix $C_{\Phi^{(2,2)}_{\alpha,\beta}}$ corresponding to the map $\Phi^{(2,2)}_{\alpha,\beta}: M_{4}{(\mathbb{C})} \rightarrow M_{4}{(\mathbb{C})}$ to construct a witness operator. Thus, we are showing the construction by taking a simple case when the domain of the map contain only $2 \otimes 2$ density matrices but one may follow the same procedure to construct the witness operator when the domain of the map contain only $d_{1} \otimes d_{2}$ density matrices.\\
The constructed witness operator can detect positive partial transpose entangled states as well as negative partial transpose entangled states that may exist in any arbitrary dimension. To start with the construction of witness operator, we first define a Hermitian operator $O_{\alpha,\beta} \in M_2(M_8)$ as
\begin{equation}
O_{\alpha,\beta} = C_{\Phi_{\alpha,\beta}} C_{\Phi_{\alpha,\beta}}^{\dagger} =
\begin{bmatrix}
A & B \\
B^{\dagger} & D \\
\end{bmatrix}
\label{operator1}
\end{equation}
where $A, B, D$ denote the $8 \times 8$ block matrices.\\
The $8 \times 8$ block matrices $A, B, D$ can further be expressed in terms of $4 \times 4$ block matrices as
\begin{equation}
A=
\begin{pmatrix}
A_{11}&A_{12}\\
A_{21}&A_{22}\\
\end{pmatrix};
B=
\begin{pmatrix}
B_{11}&B_{12}\\
B_{21}&B_{22}\\
\end{pmatrix};
D=
\begin{pmatrix}
D_{11}&D_{12}\\
D_{21}&D_{22}\\
\end{pmatrix}
\end{equation}
where the $4 \times 4$ block matrices are given by
\begin{equation}
A_{11}=
\begin{pmatrix}
\alpha^2 + \beta^2 + (\alpha + \beta)^2 & 0& 0& 0 \\
0& 2 \alpha^2 + 2 \beta^2& 0& 0\\
0& 0& 0& 0 \\
0& 0& 0& 0 \\
\end{pmatrix};\nonumber\\
\end{equation}
\begin{equation}
A_{12}=
\begin{pmatrix}
0& 2\alpha\beta& 0& 0 \\
0& 0& 0& 0 \\
0& 0& 0& 0 \\
0& 0& 0& 0 \\
\end{pmatrix}; A_{21}=A_{12}^{\dagger};\nonumber\\
\end{equation}
\begin{equation}
A_{22}=
\begin{pmatrix}
2 \alpha^2 + 2 \beta^2& 0& 0& 0 \\
0& \alpha^2 + \beta^2 + (\alpha + \beta)^2& 0& 0 \\
0& 0& 0& 0 \\
0& 0& 0& 0 \\
\end{pmatrix}
\end{equation}
\begin{equation}
B_{11}=
\begin{pmatrix}
0& 0& \alpha^2 + \beta^2 + (\alpha + \beta)^2& 0\\
0& 0& 0& 2 \alpha^2 + 2 \beta^2\\
0& 0& 0& 0\\
0& 0& 0& 0\\
\end{pmatrix};\nonumber\\
\end{equation}
\begin{equation}
B_{12}=
\begin{pmatrix}
0& 0& 0& 2\alpha\beta\\
0& 0& 0& 0\\
0& 0& 0& 0\\
0& 0& 0& 0\\
\end{pmatrix};
B_{21}=
\begin{pmatrix}
0& 0& 0& 0\\
0& 0& 2 \alpha\beta& 0\\
0& 0& 0& 0\\
0& 0& 0& 0\\
\end{pmatrix};\nonumber\\
\end{equation}
\begin{equation}
B_{22}=
\begin{pmatrix}
0& 0& 2 \alpha^2 + 2 \beta^2& 0\\
0& 0& 0& \alpha^2 + \beta^2 + (\alpha + \beta)^2\\
0& 0& 0& 0\\
0& 0& 0& 0\\
\end{pmatrix}
\end{equation}
\begin{equation}
D_{11}=
\begin{pmatrix}
0& 0& 0& 0\\
0& 0& 0& 0\\
0& 0& \alpha^2 + \beta^2 + (\alpha + \beta)^2& 0\\
0& 0& 0& 2 \alpha^2 + 2 \beta^2\\
\end{pmatrix};\nonumber\\
\end{equation}
\begin{equation}
D_{12}=
\begin{pmatrix}
0& 0& 0& 0\\
0& 0& 0& 0\\
0& 0& 0& 2\alpha\beta\\
0& 0& 0& 0\\
\end{pmatrix}; D_{21} = D_{12}^\dagger;\nonumber\\
\end{equation}
\begin{equation}
D_{22}=
\begin{pmatrix}
0& 0& 0& 0\\
0& 0& 0& 0\\
0& 0& 2 \alpha^2 + 2 \beta^2& 0\\
0& 0& 0& \alpha^2 + \beta^2 + (\alpha + \beta)^2\\
\end{pmatrix}
\end{equation}
Now, we are in a position to use the Hermitian operator $O_{\alpha,\beta}$ defined in (\ref{operator1}) for the construction of a witness operator $W_{\alpha,\beta}$ in $4\otimes4$ dimensional space to detect an entangled state lying in the same space. We define an operator $W_{\alpha,\beta}$ as
\begin{equation}
W_{\alpha,\beta}= O_{\alpha,\beta} - \gamma I_{16}
\end{equation}
where $I_{16}$ denote the identity matrix of order 16.\\
Our task is now to show that $W_{\alpha,\beta}$ is an witness operator for some suitable $\gamma$. To show this, we need to prove the following facts:\\
\textbf{(i)} $Tr(W_{\alpha,\beta}\sigma)\geq 0$ for all separable state $\sigma$.\\
\textbf{(ii)} $Tr(W_{\alpha,\beta}\rho_{e})< 0$ for at least one entangled state $\rho_{e}$.\\
To prove \textbf{(i)}, let us consider a separable state $\sigma$ lying in $4\otimes4$ dimensional space. The state $\sigma$ is given in the form as
\begin{equation}
\sigma =
\begin{pmatrix}
X & Y \\
Y^{\dagger} & Z \\
\end{pmatrix}
\end{equation}
where $X,Z\geq 0$, and $Y$ represent a $8 \times 8$ block matrices and satisfies $X\geq YZ^{-1}Y^{\dagger}$.
Since $\sigma$ represent a separable state so $\sigma^{T_{B}}=\begin{pmatrix}
X^{T} & Y^{T} \\
(Y^{\dagger})^{T} & Z^{T} \\
\end{pmatrix}$ represent a positive semi-definite matrix and thus the block matrices $X^{T}$, $Y^{T}$, $(Y^{\dagger})^{T}$ and $Z^{T}$ satisfies
the following inequalities \cite{sharma}
\begin{equation}
X^{T}\geq 0,~~ Z^{T}\geq 0,~~ X^{T}\geq Y^{T}(Z^{T})^{-1}(Y^{\dagger})^{T}
\label{semi-def1}
\end{equation}
The expectation value of the operator $W_{\alpha,\beta}$ with respect to the state $\sigma$ is given by
\begin{equation}
\langle W_{\alpha,\beta}\rangle_{\sigma}=Tr[W_{\alpha,\beta} \sigma] = Tr[O_{\alpha,\beta} \sigma] - \gamma
\label{expectation}
\end{equation}
$Tr[O_{\alpha,\beta} \sigma]$ given in (\ref{expectation}) can be calculated as
\begin{eqnarray}
Tr[O_{\alpha,\beta} \sigma] &=& Tr[AX + BY^{\dagger}] + Tr[B^{\dagger}Y + DZ] \nonumber\\
&=& Tr[AX] + Tr[DZ] + 2Tr[BY^{\dagger}]\nonumber\\
&\geq& \lambda_{min}(X) Tr[A] + \lambda_{min}(Z) Tr[D] + 2 Tr[BY^{\dagger}] \nonumber\\
&=& (\lambda_{min}(X) + \lambda_{min}(Z)) Tr[A] + 2 Tr[BY^{\dagger}] \label{expectation1}
\end{eqnarray}
The third step follows from Result-2 given in (\ref{res2}) and last step holds true since $Tr[A] = Tr[D]$.\\
Using (\ref{expectation1}) in the expression (\ref{expectation}), we get
\begin{eqnarray}
Tr[W_{\alpha,\beta} \sigma] &\geq&  (\lambda_{min}(X) + \lambda_{min}(Z)) Tr[A] + 2 Tr[BY^{\dagger}]\nonumber\\&-&\gamma
\label{wit1}
\end{eqnarray}
Choosing $\gamma=2 (Tr[BY^{\dagger}]+ Tr[A] ||Y||_2)$, the inequality (\ref{wit1}) reduces to
\begin{eqnarray}
Tr[W_{\alpha,\beta} \sigma] &\geq&  (\lambda_{min}(X) + \lambda_{min}(Z)) Tr[A] \nonumber\\&-& 2Tr[A]||Y||_2
\label{wit2}
\end{eqnarray}
Since $\lambda_{min}(X)$ and $\lambda_{min}(Z)$ are non-negative so we can apply $AM \geq GM$ on $\lambda_{min}(X)$ and $\lambda_{min}(Z)$ and thus the inequality (\ref{wit2}) further reduces to
\begin{eqnarray}
Tr[W_{\alpha,\beta} \sigma] \geq 2Tr[A] ((\lambda_{min}(X)\lambda_{min}(Z))^{\frac{1}{2}}-||Y||_2) \nonumber
\label{wit3}
\end{eqnarray}
Using the separability criterion given in (\ref{res1}) and the fact that $Tr[A]\geq 0$, we can conclude that $Tr[W_{\alpha,\beta} \sigma] \geq 0$ for any separable state $\sigma$.\\
To prove \textbf{(ii)}, let us consider a $4\otimes4$ dimensional bound entangled state $\rho^{(p,q)}$ given by \cite{akbari}\\
\begin{eqnarray}
\rho^{(p,q)} = p \sum_{i=1}^4 |\omega_i\rangle \langle\omega_i| +  q \sum_{i=5}^6 |\omega_i\rangle \langle\omega_i| \label{bes4by4}
\end{eqnarray} where $p$ and $q$ are non-negative real numbers and $4p + 2q = 1$. The pure states $\{|\omega_i\rangle\}_{i=1}^6$ are defined as follows:
\begin{eqnarray*}
&& |\omega_1\rangle = \frac{1}{\sqrt{2}} (|01\rangle + |23\rangle)\\
&& |\omega_2\rangle = \frac{1}{\sqrt{2}} (|10\rangle + |32\rangle)\\
&& |\omega_3\rangle = \frac{1}{\sqrt{2}} (|11\rangle + |22\rangle)\\
&& |\omega_4\rangle = \frac{1}{\sqrt{2}} (|00\rangle - |33\rangle)\\
&& |\omega_5\rangle = \frac{1}{2} (|03\rangle + |12\rangle) + \frac{|21\rangle }{\sqrt{2}}\\
&& |\omega_6\rangle = \frac{1}{2} (-|03\rangle + |12\rangle)+ \frac{|30\rangle }{\sqrt{2}}
\end{eqnarray*}
The state $\rho^{(p,q)}$ becomes invariant under partial transposition when $p = \frac{q}{\sqrt2}$ which implies that $\rho^{(p,q)}$ is a PPT state for $q = \frac{\sqrt{2} - 1}{2}\equiv q_{1}$ and $p = \frac{1 - 2q}{4}\equiv p_{1}$. Since $\|(\rho^{(p_{1},q_{1})})^R\|_1 = 1.08579$, which is greater than 1, so by matrix realignment criteria explained in Result-4, one can say that $\rho^{(p_{1},q_{1})}$ is a PPT entangled state.
A simple calculation shows that $\rho^{(p_{1},q_{1})}$ is detected by our witness operator $W_{\alpha,\beta}$ for all $\alpha>0, \beta > 0$. To show it explicitly, let us calculate $Tr(W_{\alpha,\beta} \rho^{(p_{1},q_{1})})$. It gives
\begin{eqnarray}
Tr(W_{\alpha,\beta} \rho^{(p_{1},q_{1})}) = -1.07107 (2\alpha^2 + \alpha\beta + 2\beta^2) \\\nonumber \quad \forall \; \alpha, \beta >0
\end{eqnarray}



Since, $W_{\alpha,\beta}$ detects the BES described by the density operator $\rho^{(p_{1},q_{1})}$, so $W_{\alpha,\beta}$ satisfies both \textbf{(i)} and \textbf{(ii)} and thus it qualifies to become a witness operator for all $\alpha>0$ and $\beta \geq 0$.
\section{Efficiency of the constructed witness operator $W_{\alpha,\beta}$}
In this section, we study the detection power of our criteria by comparing with other powerful existing entanglement detection criterion in the literature \cite{rudolph1,kchen,dv,sarbicki}. We take few examples of the family of PPTES and then show that our witness operator detect most PPTES in the family in comparison to other well known entanglement criterion such as the realignment or computable cross norm (CCNR) criterion \cite{rudolph1}, the de Vicente criterion (dV) \cite{dv} and the separability criterion based on correlation tensor (CT) given in \cite{sarbicki}.\\
\subsection{Example-1}
In this subsection, we illustrate the efficiency of our witness operator $W_{\alpha,\beta}$ using a class of $4\otimes 4$ bound entangled state which cannot be detected by realignment criteria.\\
Consider the following $4\otimes4$ PPT entangled state \cite{kye}.\\
\begin{equation}
\rho_{z,p,r}= \frac{1}{N}
\begin{pmatrix}
A_{11} & A_{12} & A_{13} & A_{14}\\
A_{12}^* & A_{22} & A_{23} & A_{24}\\
A_{13}^* & A_{23}^* & A_{33} & A_{34}\\
A_{14}^* & A_{24}^* & A_{34}^* & A_{44}\\
\end{pmatrix}
\end{equation}
where $ A_{11} =
\begin{pmatrix}
z + \overline{z} & 0 & 0 & 0\\
0 & \frac{1}{p} & 0 & 0\\
0 & 0 & p & 0\\
0 & 0 & 0 & \frac{r}{p} + r\\
\end{pmatrix}\\
A_{12} =
\begin{pmatrix}
0 & -z & 0 & 0\\
0 & 0 & 0 & 0\\
0 & 0 & 0 & -r\\
0 & 0 & 0 & 0\\
\end{pmatrix};
A_{13} =
\begin{pmatrix}
0 & 0 & -\overline{z} & 0\\
0 & 0 & 0 & -rz\\
0 & 0 & 0 & 0\\
0 & 0 & 0 & 0\\
\end{pmatrix}\\
A_{22} =
\begin{pmatrix}
p & 0 & 0 & 0\\
0 & z + \overline{z} & 0 & 0\\
0 & 0 & \frac{r}{p} + r & 0\\
0 & 0 & 0 & \frac{1}{p}\\
\end{pmatrix};
A_{24} =
\begin{pmatrix}
0 & 0 & -rz & 0\\
0 & 0 & 0 & -z\\
0 & 0 & 0 & 0\\
0 & 0 & 0 & 0\\
\end{pmatrix};\\
A_{33} =
\begin{pmatrix}
\frac{1}{p} & 0 & 0 & 0\\
0 & rp + r & 0 & 0\\
0 & 0 & z + \overline{z} & 0\\
0 & 0 & 0 & p\\
\end{pmatrix};
A_{34} =
\begin{pmatrix}
0 & -r & 0 & 0\\
0 & 0 & 0 & 0\\
0 & 0 & 0 & -\overline{z}\\
0 & 0 & 0 & 0\\
\end{pmatrix}\\
A_{44} =
\begin{pmatrix}
rp + r & 0 & 0 & 0\\
0 & p & 0 & 0\\
0 & 0 & \frac{1}{p} & 0\\
0 & 0 & 0 & z + \overline{z}\\
\end{pmatrix}; A_{14} = A_{23} = [0]_{4\times4}\\$ and $N = Tr[\rho_{z,p,r}]$.\\
$\rho_{z,p,r}$ is PPTES when $p>0$, $0<r<1$ and $z$ is a complex number such that $|z|=1$ and $\frac{-\pi}{4}<Arg(z)<\frac{\pi}{4}$.
If we take $p=z=1$ then we find that the state $\rho_{1,1,r}$ is not detected by realignment criterion, since $||R(\rho_{1,1,r})||_1 = \frac{2}{2+r} < 1$. $R(\rho_{1,1,r})$ denoting the realigned matrix of $\rho_{1,1,r}$.  \\
Now our task is to see whether the witness operator constructed here is able to detect the state described by the density operator $\rho_{1,1,r}$. To probe this fact, we calculate the expectation value of witness operator $W_{\alpha,\beta}$ for the state $\rho_{1,1,r}$. It is given by
\begin{eqnarray}
Tr[W_{\alpha,\beta}\rho_{1,1,r}]
&=& \frac{1}{2+r} ((\alpha^2 + \beta^2)(3-2\sqrt{2+2r^2})\nonumber\\ && +  \alpha \beta (1-\sqrt{2+2r^2})) \\ \nonumber
\end{eqnarray}
After simple calculation, we observed the following:
\begin{itemize}
  \item[{1.}] If $r \in (0,\frac{1}{2\sqrt{2}})$ then $Tr[W_{\alpha,\beta}\rho_{1,1,r}] < 0$ for $\frac{\alpha\beta}{\alpha^2 + \beta^2} > \frac{3-2\sqrt{2-2r^2}}{\sqrt{2+2r^2} - 1}$.
  \item[2.] If $r \in [\frac{1}{2\sqrt{2}},1)$ then $Tr[W_{\alpha,\beta}\rho_{1,1,r}] < 0$ for any $\alpha, \beta > 0$.
\end{itemize}
The above facts implies that there exist positive parameters $\alpha$ and $\beta$ so that $Tr[W_{\alpha,\beta}\rho_{1,1,r}] < 0$ in the whole range $0 < r < 1$. Thus, $\rho_{1,1,r}$ is detected by $W_{\alpha,\beta}$ for some positive values of $\alpha$ and $\beta$. Hence, our witness operator $W_{\alpha,\beta}$ detects this family of PPTES in the whole range $0 < r < 1$.\\
For comparison, let us examine other separability criteria for detection of entanglement belonging to the family $\rho_{1,1,r}$. The comparison is summarized in the table given below:\\
\begin{table}[h!]
\begin{tabular}{| p{3.7cm} | p{4.3cm} |}
\hline
\multicolumn{2}{|c|}{The state $\rho_{1,1,r},~~0<r<1$} \\
\hline
Criterion & Detection range \\
\hline
\hline
1. dV &  Does not detect \\
2. CCNR &  Does not detect \\
3. CT & Does not detect\\
4. Witness operator $W_{\alpha,\beta}$ & $0< r < 1$\\

\hline
\end{tabular}
\caption{Detection of PPTES $\rho_{1,1,r}$ in the range $0 < r < 1$}
\label{atable}
\end{table}
CT criterion does not detect entanglement in the family of states $\rho_{1,1,r}$ for $0 < r < 1$. $\rho_{1,1,r}$  remains undetected using dV and CCNR criteria.

\subsection{Example-2}
We construct a one parameter family of 2 ququart states with positive partial transpose. These states are obtained by mixing the bound entangled state $\rho_{p_1,q_1}$ described in (\ref{bes4by4}) with white noise:
 \begin{equation}
 \rho^{\lambda}_{p_1,q_1} = \lambda\rho_{p_1,q_1} + \frac{1-\lambda}{16} I_{4} \otimes I_4,~~0\leq \lambda \leq 1
 \end{equation}
 We note the following facts:\\
(i) $\rho^{\lambda}$ is invariant under partial transposition.\\
(ii) \textbf{Realignment criteria:} By matrix realignment criteria, $\rho^{\lambda}$ is entangled when $\lambda \in (0.897358,1]$.\\
(iii) \textbf{CT criteria:} To apply CT criteria, we calculate the correlation matrix $C^{can}_{\lambda}$ for $\rho^{\lambda}_{p_1,q_1}$ using generalized Gell-Mann matrix basis \cite{bert}. Using (\ref{ct}), we have
\begin{eqnarray}
&&||D_{x}^AC^{can}_{\lambda} D_{y}^B||_1 - \mathcal{N}_A (x) \mathcal{N}_B (y)\nonumber\\ \nonumber &=& \frac{1}{4} ((9 - 4\sqrt{2})\lambda + xy - \sqrt{(3 + x^{2})(3+y^2)})\\ 
&\leq& 0 \label{dvin} \\\label{bound} 
&&\text{when} \quad \lambda \leq \frac{\sqrt{(3+x^2)(3+y^2)}-xy}{9-4\sqrt{2}}
\label{cr1}
\end{eqnarray}
We find that the upper bound of $\lambda$ given in (\ref{cr1}) is tight. For $x=y$, the upper bound of $\lambda$ is found out to be $\frac{3}{9- 4\sqrt{2}}= 0.897358$.  This implies that the inequality in (\ref{dvin}) is violated for any $x=y\geq0$ when $\lambda >0.897358$ and hence PPT entangled states in this region are detected by CT criterion.\\
In particular, if $(x=y=0)$ then CT criterion reduces to dV criteria and it also able to detect the entangled state in the same region as realignment criteria and CT criterion. We further note that even for large values of $x$ and $y$, other PPTES of this family are not detected by CT criteria.\\
(iv) \textbf{Our criteria:} We now examine our criteria to detect entanglement in the family of states $\rho^{\lambda}_{p_1,q_1}$. To apply our criteria, we need to construct witness operator for the detection of entangled states in the family represented by $\rho^{\lambda}_{p_1,q_1}$. The witness operator $W_{\alpha,\beta}$ can be constructed through the prescription given in the previous section. It is given by
\begin{eqnarray}
W_{\alpha,\beta}= O_{\alpha,\beta} - \gamma I
\end{eqnarray}
where $\gamma = (-6 + 7\sqrt{2})(\alpha^2 + \beta^2) + 4(-1 + \sqrt2)\alpha\beta$.\\
The expectation value of $W_{\alpha,\beta}$ with respect to the state $\rho^\lambda_{p_1,q_1}$ is given by
\begin{equation}
Tr[W_{\alpha,\beta} \rho^\lambda_{p_1,q_1}] = \frac{1}{2} (2\alpha^2 + \alpha\beta + 2\beta^2)(1 + (11-10\sqrt{2})\lambda)
\end{equation}
Therefore, it can be easily shown that $Tr[W_{\alpha,\beta} \rho^\lambda_{p_1,q_1}] < 0$ for $\alpha,\beta > 0$ and $\lambda> \frac{1}{-11+10\sqrt{2}} \approx 0.318255$.\\
Thus our witness operator detects entanglement in the range $0.318255< \lambda \leq 1$, which is better than dV, CCNR and CT criterion.
We now summarize the results in Table-\ref{4by4}. It shows the range in which $\rho^{\lambda}_{p_1,q_1}$ is detected using different separability criteria.
\begin{table}[h!]
\begin{tabular}{| p{3.7cm} | p{4.3cm} |}
\hline
\multicolumn{2}{|c|}{The family of states $\rho^{\lambda}_{p_1,q_1}$} \\
\hline
Criterion & Detection range  \\
\hline
\hline
1. dV &  $0.897358< \lambda \leq 1$ \\
2. CCNR &  $0.897358< \lambda \leq 1$ \\
3. CT & $0.897358< \lambda \leq 1$ \\

4. Witness operator $W_{\alpha,\beta}$ & $0.318255< \lambda \leq 1$\\
\hline
\end{tabular}
\caption{Detection of BES in $\rho^{\lambda}_{p_1, q_1}$ in the range $0\leq \lambda \leq 1$}
\label{4by4}
\end{table}
\section{Conclusion}
To summarize, we have constructed a linear map which is shown to be a positive map under certain restrictions on the parameters involved in the construction of the map. To make our discussion simple, we have taken $2\otimes 2$ dimensional system and then find out the condition for which the map is positive. To investigate the completely positivity of the map, we start with the Choi matrix associated with the constructed map and shown that the Choi matrix has always at least one negative eigenvalue for the parameters with respect to which the map is positive. Thus the Choi matrix is not a positive semi-definite matrix and hence the map is not completely positive. Furthermore, we find that the Choi matrix is not Hermitian so we use the product of the Choi matrix and its conjugate transpose. The resulting product now represents a Hermitian matrix. In the next step, we take the linear combination of the obtained product and the identity matrix to construct the witness operator. The constructed witness operator not only detect NPTES but also may be used to detect PPTES. Since the Choi matrix is generated from the linear map defined in (\ref{phi}) so if we consider $d\otimes d$ dimensional system as the input of the map $\phi$ then the generated Choi matrix is of order $d^{2}$. Also since the construction of our witness operator depends on the Choi matrix so our witness operator may detect NPTES and PPTES lying $d^2 \otimes d^2$ dimensional Hilbert space.  Interestingly, we found that our witness operator is efficient in detecting several bipartite bound entangled states which were previously undetected by some well-known separability criteria such as realignment criterion. Moreover, we also compared the detection power of our witness operator with three well-known separability criteria, namely, dV criterion, CCNR criterion and the separability criteria based on correlation tensor (CT) proposed by Sarbicki et. al. \cite{sarbicki} and found that our witness operator detect more PPTES than the criteria mentioned above.

\section{Acknowledgement}
The first author Shruti Aggarwal would like to acknowledge the financial support by Council of Scientific and Industrial Research (CSIR), Government of India (File no. 08/133(0043)/2019-EMR-1).

\section{DATA AVAILABILITY STATEMENT}
Data sharing not applicable to this article as no datasets were generated or analysed during the current study

\section{Appendix: Detailed Calculation}
\example
\textbf{(i) Detection of $\rho_{1,1,r}$ using our Witness operator $W_{\alpha,\beta} $}\\
The state $\rho_{1,1,r}$ can be expressed in the block matrix form as
\begin{equation}
	\rho_{1,1,r} =
	\begin{pmatrix}
		X_r & Y_r \\
		Y_r^{\dagger} & Z_r \\
	\end{pmatrix}
\end{equation}
where $X_r, Y_r$ and $Z_r$ are $8 \times 8$ block matrices given as

\begin{equation*}
	X_r = \frac{1}{16+ 8r}
	\begin{pmatrix}
		2 & 0 & 0 & 0 & 0 & -1 &0 &0 \\
		0 & 1 & 0 & 0&0 & 0 & 0 & 0\\
		0 & 0 & 1 & 0&0 & 0 & 0 & -r\\
		0 & 0 & 0 & 2r &0 & 0 & 0 & 0\\
		0 & 0 & 0 & 0&1 & 0 & 0 & 0\\
		-1 & 0 & 0 & 0&0 & 2 & 0 & 0\\
		0 & 0 & 0 & 0&0 & 0 & 2r & 0\\
		0 & 0 & -r & 0&0 & 0 & 0 & 1\\
	\end{pmatrix}
\end{equation*}

\begin{equation*}
	Y_r = \frac{1}{16+ 8r}
	\begin{pmatrix}
		0 & 0 & -1 & 0 & 0 & 0 &0 &0 \\
		0 & 0 & 0 & -r &0 & 0 & 0 & 0\\
		0 & 0 & 0 & 0&0 & 0 & 0 & 0\\
		0 & 0 & 0 & 0&0 & 0 & 0 & 0\\
		0 & 0 & 0 & 0&0 & 0 & -r & 0\\
		0 & 0 & 0 & 0&0 & 0 & 0 & -1\\
		0 & 0 & 0 & 0&0 & 0 & 0 & 0\\
		0 & 0 & 0 & 0&0 & 0 & 0 & 0\\
	\end{pmatrix}
\end{equation*}

\begin{equation*}
	Z_r = \frac{1}{16 + 8r}
	\begin{pmatrix}
		1 & 0 & 0 & 0 & 0 & -r &0 &0 \\
		0 & 2r & 0 & 0&0 & 0 & 0 & 0\\
		0 & 0 & 2 & 0&0 & 0 & 0 & -1\\
		0 & 0 & 0 & 1&0 & 0 & 0 & 0\\
		0 & 0 & 0 & 0&2r & 0 & 0 & 0\\
		-r & 0 & 0 & 0&0 & 1 & 0 & 0\\
		0 & 0 & 0 & 0&0 & 0 & 1 & 0\\
		0 & 0 & -1 & 0&0 & 0 & 0 & 2\\
	\end{pmatrix}
\end{equation*}
$\rho_{1,1,r}$ is a PPT state since eigenvalues of $\rho_{1,1,r}^{T_B}$  are non negative.
By matrix realignment criterion, $||\rho_{1,1,r}^R||_1= \frac{2}{2+r} < 1$ when $0<r<1$ which implies that entanglement in the state $\rho_{1,1,r}$ is undetected by realignment criterion.\\
Now we use our entanglement witness operator $W_{\alpha,\beta}$ to detect this state.
\begin{eqnarray*}
	W_{\alpha,\beta}= O_{\alpha,\beta} - \gamma I
\end{eqnarray*}
where $O_{\alpha,\beta}= C_{\Phi_{\alpha,\beta}} C_{\Phi_{\alpha,\beta}}^{\dagger}$ is explicitly defined in section IV; and

\begin{eqnarray*}
	\gamma&=& 2 (Tr[BY_r^{\dagger}]+ Tr[A] ||Y_r||_2)\\
	&=&\frac{1}{2+r}((-1-r+2\sqrt{2(1+r^2)})(\alpha^2 + \beta^2)\\ &&+ ((-1 + \sqrt{2(1+r^2)})\alpha\beta)
\end{eqnarray*}

where
$$||Y_r||_2 = \frac{\sqrt{1 + r^2}}{\sqrt2 (8+4r)}$$ 
$$\\Tr[A]= 6 \alpha^2 + 6 \beta^2 + 2(\alpha+\beta)^2$$
$$Tr[BY_r^{\dagger}]= -\frac{\alpha\beta+(\alpha^2 +\beta^2)(1+r)}{2(2+r)}$$

The expectation value of $W_{\alpha,\beta}$ with respect to the state $\rho^\lambda_{p_1,q_1}$ is given by
\begin{eqnarray*}
	Tr[W_{\alpha,\beta}\rho_{1,1,r}]
	&=& \frac{1}{2+r} ((\alpha^2 + \beta^2)(3-2\sqrt{2+2r^2})\\ &&+  \alpha \beta (1-\sqrt{2+2r^2}))\\  
\end{eqnarray*}

$Tr[W_{\alpha,\beta}\rho_{1,1,r}]<0$ when $\frac{\alpha\beta}{\alpha^2 + \beta^2} > \frac{3-2\sqrt{2-2r^2}}{\sqrt{2+2r^2} - 1} $
where $\alpha,\beta >0$ and $0<r<1$.\\


 Thus, $\rho_{1,1,r}$ is detected by $W_{\alpha,\beta}$ for some positive values of $\alpha$ and $\beta$. Hence, our witness operator $W_{\alpha,\beta}$ detects this family of PPTES in the whole range $0 < r < 1$.\\
\textbf{(ii) CT criterion:}
The correlation matrix for the state $C^{can}_r$ for the state $\rho_{1,1,r}$ is 
\begin{equation}
	C^{can}_r =
	\begin{pmatrix}
		C_{11} & C_{12} \\
		C_{21} & C_{22} \\
	\end{pmatrix}
\end{equation}
where
\begin{eqnarray*}
	C_{11} =
	\begin{pmatrix}
		\frac{1}{4} & 0 &0 &	0 & 0 &0 &	0 & 0\\
		0 & \frac{-1}{8(2+r)} &0 &	0 & 0 &0 &	\frac{-r}{8(2+r)} & 0\\
		0 & 0 &\frac{-1}{8(2+r)} &	0 & 0 &\frac{-r}{8(2+r)} &	0 & 0\\
		0 & 0 &0 &	0 & 0 &0 &	0 & 0\\
		0 & 0 &0 &	0 & 0 &0 &	0 & 0\\
		0 & 0 &\frac{-r}{8(2+r)} &	0 & 0 &\frac{-1}{8(2+r)} &	0 & 0\\
		0 & \frac{-r}{8(2+r)} &0 &	0 & 0 &0 &	\frac{-1}{8(2+r)} & 0\\
		0 & 0 &0 &	0 & 0 &0 &	0 & \frac{1}{8(2+r)}\\
	\end{pmatrix};
\end{eqnarray*}

\begin{eqnarray*}
	C_{12} =
	\begin{pmatrix}
		0 & 0 &0 &	0 & 0 &0 &	0 & 0\\
		0 & 0 &0 &	0 & 0 &0 &	0 & 0\\
		0 & 0 &0 &	0 & 0 &0 &	0 & 0\\
		0 & 0 &0 &	0 & 0 &0 &	0 & 0\\
		0 & 0 &0 &	0 & 0 &0 &	0 & 0\\
		0 & 0 &0 &	0 & 0 &0 &	0 & 0\\
		0 & 0 &0 &	0 & 0 &0 &	0 & 0\\
		0 & 0 &0 &	0 & \frac{r}{8(2+r)} &0 &	0 & 0\\
	\end{pmatrix};
\end{eqnarray*}

\begin{eqnarray}
	C_{22}=
	\begin{pmatrix}
		C_{22}^{(11)} & C_{22}^{(12)}\\
		C_{22}^{(21)} & C_{22}^{(22)}
	\end{pmatrix}
\end{eqnarray}
where
\begin{eqnarray*}
		C_{22}^{(11)}=
	\begin{pmatrix}
		\frac{1}{8(2+r)} & 0 &0 &	\frac{r}{8(2+r)} \\
			0 & 0 &0 &	0\\
				0 & 0 &0 &	0\\
		\frac{r}{8(2+r)} & 0 &0 &	\frac{1}{8(2+r)}
	\end{pmatrix}
\end{eqnarray*}

$C_{22}^{(12)}$ and $C_{22}^{(21)}$ are zero matrices;

\begin{eqnarray*}
	C_{22}^{(22)}=
	\begin{pmatrix}
	\frac{1}{8(2+r)} &0 &	0 & 0\\
	0 & \frac{1}{8(2+r)} &	\frac{-1+2r}{8\sqrt{3}(2+r)} & \frac{1-2r}{4\sqrt{6}(2+r)}\\
	0 &\frac{-1+2r}{8\sqrt{3}(2+r)} &	\frac{5-4r}{48+24r} & \frac{1-2r}{12\sqrt{2}(2+r)}\\
	0& \frac{1-2r}{4\sqrt{6}(2+r)}  &	\frac{1-2r}{12\sqrt{2}(2+r)} & \frac{2-r}{12(2+r)}\\
	\end{pmatrix}
\end{eqnarray*}

 and $C_{21} = C_{12}^{\dagger}$
\begin{eqnarray*}
	||D_{x}^A C^{can}_{r} D_{y}^B||_1 &-& \mathcal{N}_A (x) \mathcal{N}_B (y)\\ &=& \frac{1}{4} (-1 + \frac{8}{2+r} + xy - \sqrt{(3+x^2)(3+y^2)})
	\\ &\leq& 0 \;\; \forall\; x,y\geq 0 \; \text{and}\; r \in (0, 1)
	\label{cteg1}
\end{eqnarray*}
Hence, the state $\rho_{1,1,r}$ is not detected by CT criterion for any value of $r \in (0,1)$.\\
\textbf{Few Particular cases:}\\
For $(x,y)=(0,0)$ and $(1,1)$ we have $$||D_{x}^A C^{can}_{r} D_{y}^B||_1 - \mathcal{N}_A (x) \mathcal{N}_B (y) = \frac{-4r}{2+r} <0$$
which implies dV criterion and CCNR criterion fails to detect PPT entangled states in the family $\rho_{1,1,r}$

\example

\textbf{(i) Detection of $\rho^\lambda_{p_1,q_1}$ using our Witness operator $W_{\alpha,\beta}: $}\\
The state $\rho^\lambda_{p_1,q_1}$ can be expressed in the block matrix form as
\begin{equation}
	\rho^\lambda_{p_1,q_1} =
	\begin{pmatrix}
		X_{\lambda} & Y_{\lambda} \\
		Y_{\lambda}^{\dagger} & Z_{\lambda} \\
	\end{pmatrix}
\end{equation}
where $X_{\lambda}, Y_{\lambda}$ and $Z_{\lambda}$ are $8 \times 8$ block matrices given as

\begin{eqnarray*}
	X_{\lambda}= diag\{\frac{(1+(3-2\sqrt{2})\lambda)}{16}, \frac{(1+(3-2\sqrt{2})\lambda)}{16} , \\ \frac{1-\lambda}{16}, \frac{(1 +(-5 + 4\sqrt{2})\lambda)}{16}, \frac{(1+(3-2\sqrt{2})\lambda)}{16} ,\\ \frac{(1+(3-2\sqrt{2})\lambda)}{16}, \frac{(1 +(-5 + 4\sqrt{2})\lambda)}{16}, \frac{1-\lambda}{16}\}
\end{eqnarray*}

\begin{eqnarray}
	Y_{\lambda}=
	\begin{pmatrix}
		Y_{\lambda}^{(11)}&&Y_{\lambda}^{(12)}\\
		Y_{\lambda}^{(21)}&&Y_{\lambda}^{(22)}
	\end{pmatrix}
\end{eqnarray}
where
\begin{equation*}
	Y_{\lambda}^{(11)}=
	\begin{pmatrix}
		0 & 0 & 0 &0 \\
		0 & 0 & 0 &\frac{1}{8}(2 - \sqrt{2})\lambda \\
		0 & 0 & 0 &0 \\
		0 & \frac{(-1+\sqrt2)}{4\sqrt{2}}\lambda & 0 &0
	\end{pmatrix}
\end{equation*}

\begin{equation*}
	Y_{\lambda}^{(12)}=
	\begin{pmatrix}
		0 & 0 & 0 &\frac{1}{8}(-2 + \sqrt{2})\lambda\\
		0 & 0 & 0 &0\\
		-\frac{(-1+\sqrt2)}{4\sqrt{2}}\lambda & 0 &0 & 0
	\end{pmatrix}
\end{equation*}

\begin{equation*}
	Y_{\lambda}^{(21)}=
	\begin{pmatrix}
		0 & 0 & 0 &0\\
		0 & 0 & \frac{1}{8}(2 - \sqrt{2})\lambda & 0\\
		0 & \frac{(-1+\sqrt2)}{4\sqrt{2}}\lambda & 0 & 0 \\
		0 & 0 & 0 &0 
	\end{pmatrix}
\end{equation*}

\begin{equation*}
	Y_{\lambda}^{(22)}=
	\begin{pmatrix}
		0 & 0 &\frac{1}{8}(2 - \sqrt{2})\lambda & 0\\
		0 & 0 & 0 &0\\
		\frac{(-1+\sqrt2)}{4\sqrt{2}}\lambda & 0 &0 & 0\\
		0 & 0 & 0 &0
	\end{pmatrix}
\end{equation*}

\begin{eqnarray*}
	Z_{\lambda}= diag\{\frac{1-\lambda}{16}, \frac{(1 +(-5 + 4\sqrt{2})\lambda)}{16}, \frac{(1+(3-2\sqrt{2})\lambda)}{16},\\ \frac{(1+(3-2\sqrt{2})\lambda)}{16},
	\frac{(1 +(-5 + 4\sqrt{2})\lambda)}{16}, \frac{1-\lambda}{16}, \\ \frac{(1+(3-2\sqrt{2})\lambda)}{16},\frac{(1+(3-2\sqrt{2})\lambda)}{16} \}
\end{eqnarray*}
The witness operator $W_{\alpha,\beta}$ is given by
\begin{eqnarray}
	W_{\alpha,\beta}= O_{\alpha,\beta} - \gamma I
\end{eqnarray}
where $O_{\alpha,\beta}= C_{\Phi_{\alpha,\beta}} C_{\Phi_{\alpha,\beta}}^{\dagger}$ is explicitly defined in section IV, and the parameter $\gamma$ can be calculated as
\begin{eqnarray}
	\gamma&=& 2 (Tr[BY_{\lambda}^{\dagger}]+ Tr[A] ||Y_{\lambda}||_2) \nonumber\\
	&=&((-6 + 7\sqrt{2})(\alpha^2 + \beta^2) + 4(-1 + \sqrt2)\alpha\beta)\lambda
\end{eqnarray}

where
$$||Y_{\lambda}||_2 = \frac{\sqrt{3 - 2 \sqrt{2}}}{2}  \lambda$$ 
$$\\Tr[A]= 6 \alpha^2 + 6 \beta^2 + 2(\alpha+\beta)^2$$
$$Tr[BY_{\lambda}^{\dagger}]= \frac{1}{2} (2-\sqrt2)(a^2 +b^2) \lambda$$

The expectation value of $W_{\alpha,\beta}$ with respect to the state $\rho^\lambda_{p_1,q_1}$ is given by
\begin{equation}
	Tr[W_{\alpha,\beta} \rho^\lambda_{p_1,q_1}] = \frac{1}{2} (2\alpha^2 + \alpha\beta + 2\beta^2)(1 + (11-10\sqrt{2})\lambda)
\end{equation}
Therefore, it can be easily shown that $Tr[W_{\alpha,\beta} \rho^\lambda_{p_1,q_1}] < 0$ for $\alpha,\beta > 0$ and $\lambda> \frac{1}{-11+10\sqrt{2}} \approx 0.318255$.\\
Thus our witness operator detects entanglement in the range (0.318255, 1] of $\lambda$.

\textbf{(ii) Detection of $\rho^{\lambda}_{p_1,q_1}$ by CT criterion:}
First we calculate the correlation matrix $C^{can}_{\lambda}$ for the state $\rho^{\lambda}_{p_1,q_1}$ using generalized Gell-Mann matrix (GGM) basis consisting of six symmetric GGM $\{G_i\}_{i=1}^{6}$; six antisymmetric GGM $\{G_i\}_{i=7}^{12}$ and three diagonal GGM $\{G_{13},G_{14}, G_{15}\}$.\\  $C^{can}_{\lambda}$ is $16 \times 16$ matrix with entries $C_{a,b} = \langle G_a \otimes G_b \rangle_{\rho^{\lambda}_{p_1,q_1}} $  where

\begin{eqnarray*}
	C_{0,0}&=& \langle G_0 \otimes G_0 \rangle_{\rho^{\lambda}_{p_1,q_1}}= \frac{1}{4}\\
	C_{2,5}&=& \langle G_2 \otimes G_5 \rangle_{\rho^{\lambda}_{p_1,q_1}}= \frac{1}{4}(2-\sqrt{2})\lambda\\	
	C_{3,3}&=& \langle G_3 \otimes G_3 \rangle_{\rho^{\lambda}_{p_1,q_1}}= \frac{1}{4}(-2+\sqrt{2})\lambda\\
	C_{4,4}&=& \langle G_4 \otimes G_4 \rangle_{\rho^{\lambda}_{p_1,q_1}}= \frac{1}{4}(2-\sqrt{2})\lambda\\	
	C_{13,14}&=& \langle G_{13} \otimes G_{14} \rangle_{\rho^{\lambda}_{p_1,q_1}}=\frac{1}{4\sqrt{3}}(-1+\sqrt{2})\lambda\\	
	C_{13,15}&=& \langle G_{13} \otimes G_{15} \rangle_{\rho^{\lambda}_{p_1,q_1}}=\frac{1}{2\sqrt{6}}(1-\sqrt{2})\lambda\\	
	C_{14,14}&=& \langle G_{14} \otimes G_{14} \rangle_{\rho^{\lambda}_{p_1,q_1}}=\frac{1}{6}(3-2\sqrt{2})\lambda\\	
	C_{14,15}&=& \langle G_{14} \otimes G_{15} \rangle_{\rho^{\lambda}_{p_1,q_1}}=\frac{1}{12}(-4+3\sqrt{2})\lambda\\
	C_{15,15}&=& \langle G_{15} \otimes G_{15} \rangle_{\rho^{\lambda}_{p_1,q_1}}=\frac{1}{12}(3-2\sqrt{2})\lambda	
\end{eqnarray*}

$C^{can}_{\lambda}$ is symmetric with $C_{2,5}=C_{5,2}$, $C_{13, 14}=C_{14,13}$, $C_{13,15}=C_{15,13}$ and $C_{14,15}=C_{15,14}$. The rest of the entries of $C^{can}_{\lambda}$ are zero.\\

CT criterion states that if $\rho^{\lambda}_{p_1,q_1}$ is separable, then
\begin{equation}
	||D_{x}^A C^{can}_{\lambda} D_{y}^B||_1 - \mathcal{N}_A (x) \mathcal{N}_B (y) \leq 0 \label{ct}
\end{equation}
where $D_x =$ diag($x, 1, 1, . . ., 1$), $D_y =$ diag($y, 1, 1, . . ., 1$),
$\mathcal{N}_A (x)= \sqrt{\frac{3 + x^2}{4}}$, $\mathcal{N}_B (y)= \sqrt{\frac{3 + y^2}{4}}$, $x,y \geq 0$.\\
The LHS of (\ref{ct}) can be calculated as
\begin{eqnarray*}
	||D_{x}^A C^{can}_{\lambda} D_{y}^B||_1 &-& \mathcal{N}_A (x) \mathcal{N}_B (y) = \\ && \frac{1}{4} ((9 - 4\sqrt{2})\lambda + xy - \sqrt{(3 + x^{2})(3+y^2)})
	\label{cr1}
\end{eqnarray*}
Therefore, equation (\ref{ct}) is satisfied $\forall\; x,y\geq 0 \; \text{ and when}\; \lambda \leq \frac{\sqrt{(3+x^2)(3+y^2)}-xy}{9-4\sqrt{2}}$. For example, for $(x,y) = (\frac{1}{16},\frac{1}{32})$, inequality (\ref{ct}) is violated when $\lambda \geq \frac{3681}{4096} \approx 0.898682$ which implies entangled states with $\lambda \in [0.898682,1]$ are detected by CT-criterion.\\

Also for the case when $x=y$, the LHS of (\ref{ct}) can be re-expressed as
\begin{eqnarray}
	||D_{x}^A C^{can}_{\lambda} D_{x}^B||_1 - \mathcal{N}_A (x) \mathcal{N}_B (x) = \frac{1}{4} ((9 - 4\sqrt{2})\lambda -3)\nonumber \\ 
	\label{ct3}
\end{eqnarray}
We can now observe that the expression given in (\ref{ct3}) is independent of $x=y$. Therefore, CT separability criterion is violated for any $x=y\geq 0$ and when $\lambda > \frac{3}{9- 4\sqrt{2}} \approx 0.897358$. Thus CT criterion detect all entangled states $\rho^{\lambda}_{p_1,q_1}$ where $\lambda \in (0.897358,1] $ and for any $x = y$.\\
Thus, when $0.897358 < \lambda \leq 1$, there exists $(x, y)$ for which the inequality given in (\ref{ct}) is violated. Thus the states described by the density operator $\rho^{\lambda}_{p_1,q_1}$ where $\lambda \in (0.897358,1] $ are entangled states and detected by CT criterion. \\

Now since Sarbicki et. al  have mentioned that $(x, y) = (1, 1)$ reproduces CCNR criterion and $(x, y) = (0, 0)$ reproduces dV criterion so using above argument for $x=y$, we conclude that these three criteria detects entanglement in the range $0.897358< \lambda \leq 1$. Hence our criterion detects better than dV, CCNR and CT criteria.

\end{document}